# Expanding the horizon of automated metamaterials discovery via quantum annealing

**Koki Kitai[1], Jiang Guo[2], Shenghong Ju[2,3], Shu Tanaka[4,5], Koji Tsuda*[1,3,6], Junichiro Shiomi*[2,3,6], and Ryo Tamura*[1,3,6,7]**

[1]Graduate School of Frontier Sciences, The University of Tokyo, Chiba 277-8568, Japan

[2]Department of Mechanical Engineering, The University of Tokyo, Tokyo 113-8654, Japan

[3]Research and Services Division of Materials Data and Integrated System, National Institute for Materials Science, Ibaraki 305-0047, Japan

[4]Green Computing Systems Research Organization, Waseda University, Tokyo 162-0042, Japan

[5]JST, PRESTO, Saitama 332-0012, Japan

[6]RIKEN Center for Advanced Intelligence Project, Tokyo 103-0027, Japan

[7]International Center for Materials Nanoarchitectonics, National Institute for Materials Science, Ibaraki 305-0044, Japan

## Abstract

Complexity of materials designed by machine learning is currently limited by the inefficiency of classical computers. We show how quantum annealing can be incorporated into automated materials discovery and conduct a proof-of-principle study on designing complex thermofunctional metamaterials consisting of $SiO_2$, SiC, and Poly(methyl methacrylate). Empirical computing time of our quantum-classical hybrid algorithm involving a factorization machine, a rigorous coupled wave analysis, and a D-Wave 2000Q quantum annealer was insensitive to the problem size, while a classical counterpart experienced rapid increase. Our method was used to design complex structures of wavelength selective radiators showing much better concordance with the thermal atmospheric transparency window in comparison to existing human-designed alternatives. Our result shows that quantum annealing provides scientists gigantic computational power that may change how materials are designed.

## Introduction

Further evolution in materials that control energy carriers, such as photon, electrons, and phonons, is a condition to realize sustainable industry and society. The key is to manipulate transport properties at the scale of characteristic length of energy carriers. Over the last decades, advances in top-down fabrications and bottom-up syntheses, together with atomistic and spectroscopic characterizations, have given us access to a nearly unlimited exploration of structures with enhanced energy transport characteristics. This situation has resulted in a number of breakthroughs in photovoltaics[1], thermal radiators[2,3], batteries[4], thermoelectrics[5], and others. Here, metamaterials are a representative case where the artificial structures produced inside a material give rise to extraordinary properties.

Now, the above successes open up a new problem that there are too many degrees of freedom in the structures to explore. Exploration among single crystals is exhausting when considering compounds, but the number of candidates becomes truly massive when extending it to include composites in a broad sense with nanoscale inhomogeneity in the composition or simply "nanostructures". Yet, some materials with such a compositional inhomogeneity exhibit superior properties over their counterparts with ordered structures, particularly in the case when the structures are smaller than the coherence length of energy carriers as shown for photons[6], phonons[7], electrons[8], and magnons[9]. Therefore, for further evolution of materials in energy technology, overcoming the challenge of massive candidates is crucial. Hence, automated materials discovery is imperative.

Automated materials discovery based on black-box optimization is an iterative process to select one candidate from a massive number of candidates (i.e., design space) and recommend it for experimental investigations[10–12]. From existing materials properties data, machine learning predicts the properties of unobserved candidates and defines an acquisition function in design space. A global optimization problem in design space is solved with respect to this acquisition function. The best candidate is selected as the next candidate material for experiments. Using the observed properties of the selected candidate, the machine learning model is updated and defines a different acquisition function for the next iteration. Repeating this procedure with the aid of the machine learning should drastically reduce a number of experimental investigations to design materials with the desired properties. There are two barriers, *statistical* and *computational*, that hinder the application of automated materials discovery to complex materials design. A statistical barrier refers to difficulty predicting properties of materials with a limited number of training data by machine learning. A computational barrier is related to the well-known hardness of global optimization defined by the acquisition function. When a fast simulator, which calculates the properties of the target materials, is used to replace the experiment, the statistical barrier is reduced so that it is not as overwhelming as the computational one, as exemplified in several recent studies[10,12]. To circumvent the difficulty with global optimization, heuristic methods such as local searches, tree searches,[13] and genetic algorithms[14] are used instead of the exhaustive search. These methods find only local minima and the quality of optimization is not high.

To overcome this computational barrier, we propose a quantum-classical hybrid algorithm employing a D-Wave quantum annealer[15]. This computer based on quantum annealing (QA)[16–20] accurately solve a particular type of combinatorial optimization called quadratic unconstrained binary optimization (QUBO) with startling speed[21]. So far, diverse applications have been reported in, e.g., biological science[22,23], machine learning[24–26], and IoT[27–29]. Figure 1 depicts the schematic procedure of our quantum-classical hybrid algorithm. Our algorithm is capable of solving a black-box optimization problem over binary variables representing a structure of materials. First, a factorization machine (FM)[30] is trained with available data to model the material's property of interest. Selection of a new candidate with respect to an acquisition function based on a trained FM boils down to a QUBO and solved by the quantum annealer. The properties of the new candidate are obtained by an atomistic simulation. In the next step, a new point is added to training data, and the FM is retrained. We call this algorithm FMQA later on.

As a proof-of-principle, we apply the method to design the metamaterial with tailored thermal radiation spectrum. Tailoring the thermal radiation is fundamentally important because every material emits and absorbs thermal radiation. From an engineering viewpoint, wavelength-selective thermal radiation (photons) is needed for a broad range of applications. For instance, engineered thermal-emission leads to high-efficiency thermophotovoltaics[31,32], incandescent light source[33], biosensing[34,35], microbolometers[36,37], imaging[38], and drying furnace[39]. Another application that is recently attracting a great deal of attention, in response to concerns of global warming and energy crises, is the radiative sky cooling that utilizes the untapped 3 K cold space as the heat sink. Previous designs on radiative cooling which focus on simultaneously blocking solar energy (0.4–4 μm) and maximizing thermal radiation loss (> 4 μm) to the surroundings have been experimentally proved working successful in dry and clear weather. However, this design strategy does not work so effectively in hot and humid areas, because the atmospheric window (8–13 μm) which allows thermal radiation to directly transmit to the outer space becomes less transparent and a large part of the downward radiation that falls outside the window will be absorbed by the radiator. Therefore, the radiator that only emits thermal radiation to the outer space through the transparency window is desirable. Such radiator can help to maximize the outgoing radiative cooling power while minimizing the radiative energy absorption from the ambient. In this study, by using our proposed optimization method, we design such radiator with a wavelength selectivity higher than the ones designed by humans.

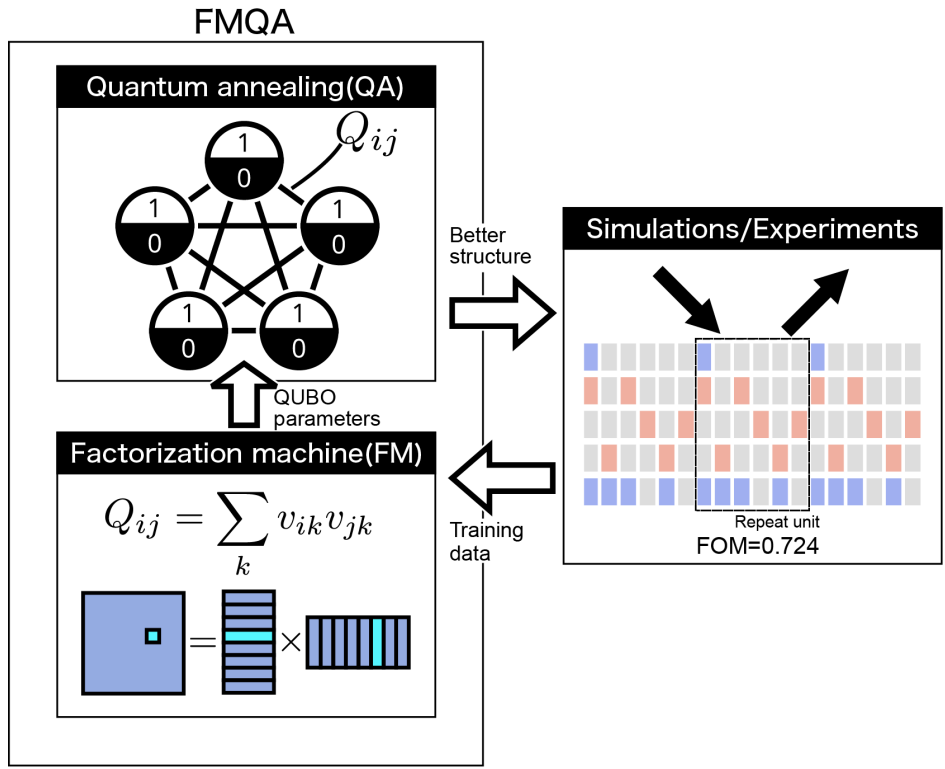

**Figure 1.** Procedure of our automated materials discovery using a factorization machine (FM) for learning and a quantum annealer (QA) for selection. Target property is the figure-of-merit (FOM) for the radiative sky cooling, which is evaluated by the rigorous coupled wave analysis (RCWA). This simulation part can be converted by other simulation methods or experiments, depending on the target properties.

# Results

## Target metamaterials

As discussed in previous studies[40,41], a metamaterial which only emits or absorbs the thermal radiation within the transparency window of the atmosphere (8–13 μm), is preferable for radiative sky cooling. Various material structures have been proposed to match the spectrum of the atmospheric window, including planar multilayer structures[2,42], patterned meta-surface structures[43–46], and polymer doped with nano-particles[47–49]. Most of these structures lack a high emittance over the whole span of the atmospheric window, and the cooling capability is insufficient. Our design employed $SiO_2$ and SiC to achieve this stringent spectral selectivity. The dielectric functions of $SiO_2$ and SiC indicate that they have phonon–polariton resonances positioned at 9.7 μm and 12.5 μm, respectively. Moreover, both have very small extinction coefficients in the solar energy wavelength band, which means the absorption of solar energy will be suppressed, thus the two materials are selected in the design.

Inspired by the previous research[50], the target metamaterial structure was comprised of $SiO_2$ and SiC wires placed in the Poly(methyl methacrylate) (PMMA), which has negligible absorption in the visible to far-infrared range. As shown in Fig. 2, each wire was arranged along the $y$-axis, and light was incident from the top layer. The periodic boundary condition is applied in the $x$ directions, i.e. the structure is repeated along the $x$ axis. The rigorous coupled wave analysis (RCWA) calculation solves a two dimensional ($x$-$z$) problem and thus the wire is assumed to be infinitely long without any variation in the cross section in the $y$ direction. For the structural optimization, the $x$-$z$ plane is uniformly discretized into square units with 1μm side lengths that are either $SiO_2$, SiC, or PMMA. The numbers of meshes along the $z$ and $x$ directions are defined as $L$ and $C$, respectively. One constraint we adopted in this study is that we only allow either $SiO_2$ or SiC in the same layer. The constraint and the unit size, which corresponds to the minimum wire size, were determined considering the possibility for fabrication in the future using photolithography. The unit size 1 μm, through pre-trial simulations, was also confirmed to give enough resolution to the optimization problem. For these structures, we calculated emissivity properties based on RCWA and evaluated the FOM for radiative cooling. In our optimization, the emittance property under p-polarized incidence (TM wave) for polar angle $\theta = 0º$ was targeted.

To utilize the quantum annealer for automated materials discovery, the structure of metamaterials should be encoded into the binary variables. Initially, the configuration of wired materials ($SiO_2$ or SiC) and PMMA was determined by $L \times C$ bits (Fig. 2). Then additional bits, which express the arrangement of wired materials in each layer, were prepared. Consequently, using $L \times (C + 1)$ bits, the structure of a metamaterial is well-defined.

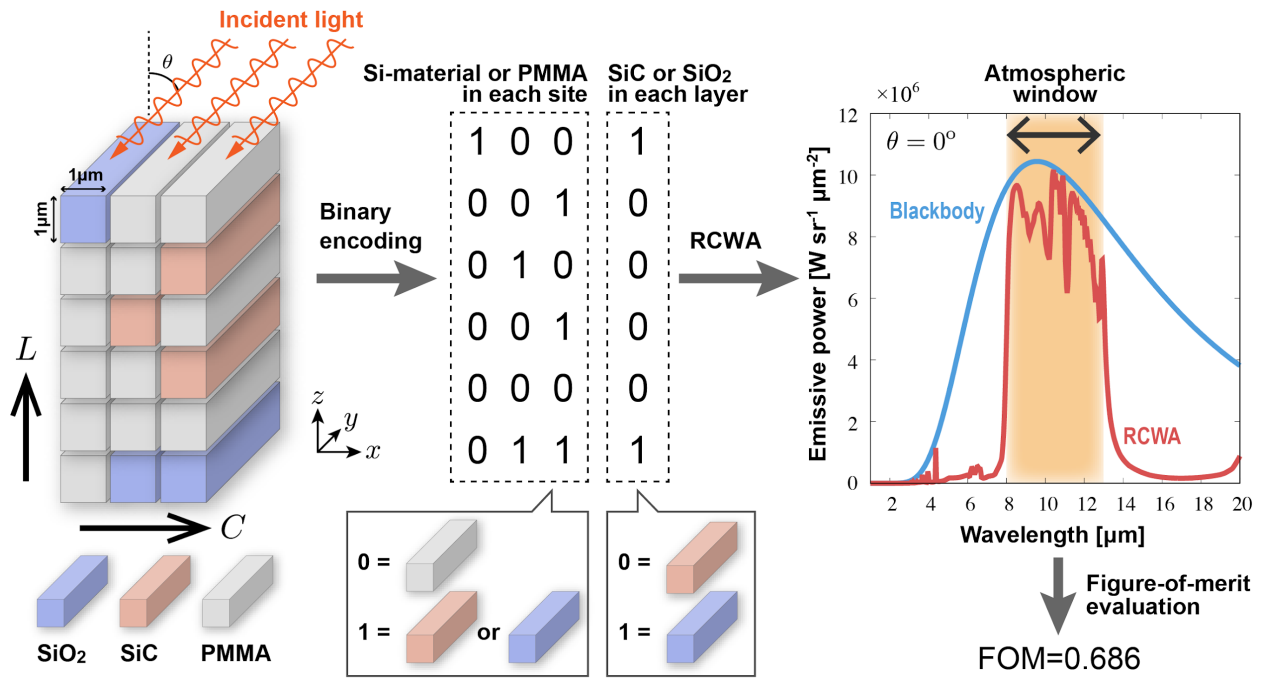

**Figure 2.** Example of the target metamaterial structure for $L = 6$ and $C = 3$, the binary variables expressing it, and the emissive power calculated by RCWA.

## Performance of FMQA

To clarify the usefulness of an FM as a regression model for our metamaterial design, we compared the performances using an FM, a Gaussian process (GP), and a random search to find the best structure for the $L = 4$ and $C = 3$ case. Since the number of candidate material structures is $2^{16} = 65,536$, we evaluated all FOMs by RCWA to identify the best structure. Furthermore, an exhaustive search of the acquisition function on the classical computers to select the next candidate material was not time consuming. Thus, a quantum annealer was not necessary in the selection part. We performed the iteration loop in automated materials discovery where the acquisition function was defined by the negative FOM predicted by the FM.

Figure 3 (a) shows the best FOM as a function of the number of calculated structures (iterations of the cycle depicted in Fig. 1) by each method. Here, 16 optimization runs with different initial choices were performed, and the FOMs were averaged out. Furthermore, the first 50 structures, which were randomly selected as the initial data, are common in all three methods. The regression results were used from step 51 for the FM and the GP cases.

The random search exhibits the worst results. Surprisingly, the FM result is better than that by the GP and the FM found the best structure within only 300 iterations. Thus, FM is suitable for the regression model of our target. This result say that at least, the learning by FM is more useful than a random search to discover the metamaterial structures with a high FOM within a small number of simulations. This result is promising for our metamaterial design.

Next, we considered the problem with $L = 6$ and $C = 3$ by FMQA. Since the candidate number was $2^{24} = 16,777,216$, an evaluation of all FOMs predicted by an FM to select the next candidate could not be performed on classical computers. Thus, it was the turn of FMQA using the D-Wave quantum annealer. Figure 3 (b) plots the best FOM by FMQA and a random search as functions of the iteration number. The average values from 16 independent runs with 50 different initial structures were plotted. FMQA can reduce the number of simulations to find the metamaterial structure with a better FOM. Hence, it is a quite useful tool to design new metamaterials. Supplementary Note A compares the effectiveness of the D-Wave quantum annealer for FMQA to the local search method.

Because the computing time to perform FMQA is important, we compared the computing time of automated materials discovery based on an FM when the selection part is conducted by the D-Wave quantum annealer (quantum-classical hybrid) to an exhaustive search by the classical computer with an Intel Xeon E5-2690 v3 @ 2.6GHz (only classical). Figure 3 (c) plots the computing time to perform 500 iterations as a function of the problem size (number of encoding bits), where the selection time, learning time by an FM, and simulation time by RCWA are separately illustrated. The target structure was the $L = 3$ case with various $C = 2, 3, 4,$ and 5, which were encoded using 9, 12, 15, and 18 bits, respectively. The empirical computing time is summarized as Supplementary Table I.

Since the learning and simulation were conducted on the same classical computer, these parts required about the same time in both cases. The selection time drastically differs, and overwhelming time reduction is succeeded by using the quantum annealer. Of course, we confirmed that both methods provide the structure with the highest FOM within only 500 iterations, and the number of simulations can be reduced to find the best one. Note that if FOMs of all structures are evaluated by the RCWA simulations for 18 bit case to identify the best structure, over ten times longer time is needed against the only classical case. Interestingly, although RCWA is a relatively high-speed simulation method (one calculation took one minute at most), the most time-consuming part becomes the simulation time when FMQA is performed. Thus, we can solve the hard *computational* barrier in the automated materials discovery with the aid of the quantum annealer.

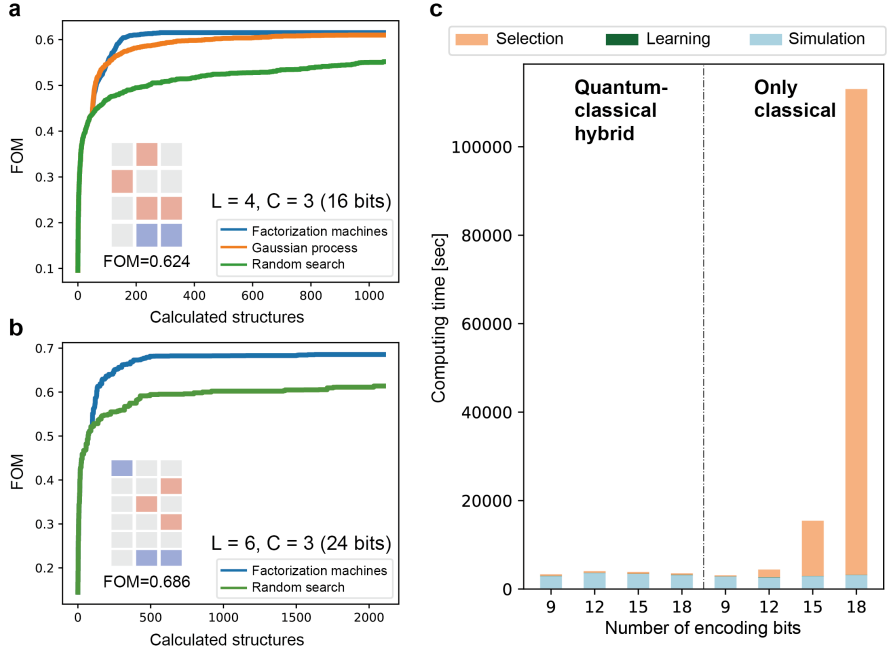

**Figure 3.** (a) Dependence of the best FOM on the number of calculated structures (iterations) by automated materials discovery using an FM, a Gaussian process, and a random search for the $L = 4$ and $C = 3$ case. Inset is the optimum structure and its FOM. Blue, red, and gray squares denote $SiO_2$, SiC, and PMMA, respectively. (b) Best FOM by FMQA using the D-Wave quantum annealer and the random search for the $L = 6$ and $C = 3$ case. Inset is the found structure with the best FOM by FMQA. (c) Computing time to perform 500 iterations of automated materials discovery using the quantum annealer (quantum-classical hybrid) and an exhaustive search on a classical computer with an Intel Xeon E5-2690 v3 @ 2.6GHz (only classical) for the selection part. Learnings and simulations are performed by the same classical computer.

## Optimum metamaterial structure search by FMQA

We searched the optimum structure of the metamaterial for radiative cooling. Varying the number of layers and columns of the target structure should help elucidate a trend to achieve a high FOM. Starting from the previous setting ($L = 6$ and $C = 3$), we initially changed the number of layers $L$. The range of $L$ was varied from 3 to 9, and 2,000 iterations were conducted for each optimization run. Figure 4 (a) plots the best FOMs as a function of the number of the calculated structures for various number of layers. The structure with five layers exhibits the highest FOM. Figure 4 (b) shows the found structure with the best FOM for each $L$. Interestingly, when the value of $L$ is more than 6, some layers only contain PMMA. If these layers are removed, the FOM decreases. Consequently, the existence of the PMMA layer plays an important role to improve the FOM for thick metamaterials.

Next we changed the number of columns $C$ in the target structures while fixing the number of layers to five. Figure 4 (c) shows the best FOM depending on the iteration number. Larger FOMs appear for $C = 4$ and 6 cases. Increasing the number of columns over seven does not yield a better FOM. Since larger cases are multiple of smaller ones with a commensurate period, a metamaterial structure for $C = 8$ should have a similar FOM as the $C = 4$ case. Thus, 2,000 samplings are too small to find the optimum structure with a higher FOM due to massive number of candidates for $C = 8$. In fact, the change in the FOMs with the number of calculated structures gradually increases for larger systems (see Fig. 4 (c)). Consequently, searches should be continued to find the best structure for larger systems.

Figure 4 (b) summarizes the metamaterial structures with a high FOM found by FMQA for various $L$ and $C$ values. The structure for the $L = 5$ and $C = 6$ case has the highest FOM, and its value is 0.724 for the radiative cooling. The emissive power of this structure is shown in Fig. 4 (d), and the large emittances fall into the atmospheric window. Note that the maximum FOM of a multilayer structure constructed by $SiO_2$, SiC, and PMMA is 0.250 for five layers. (For the detail structure, see Supplementary Note B.) This means that our designed metamaterial structure is essential to obtain high FOM for radiative cooling.

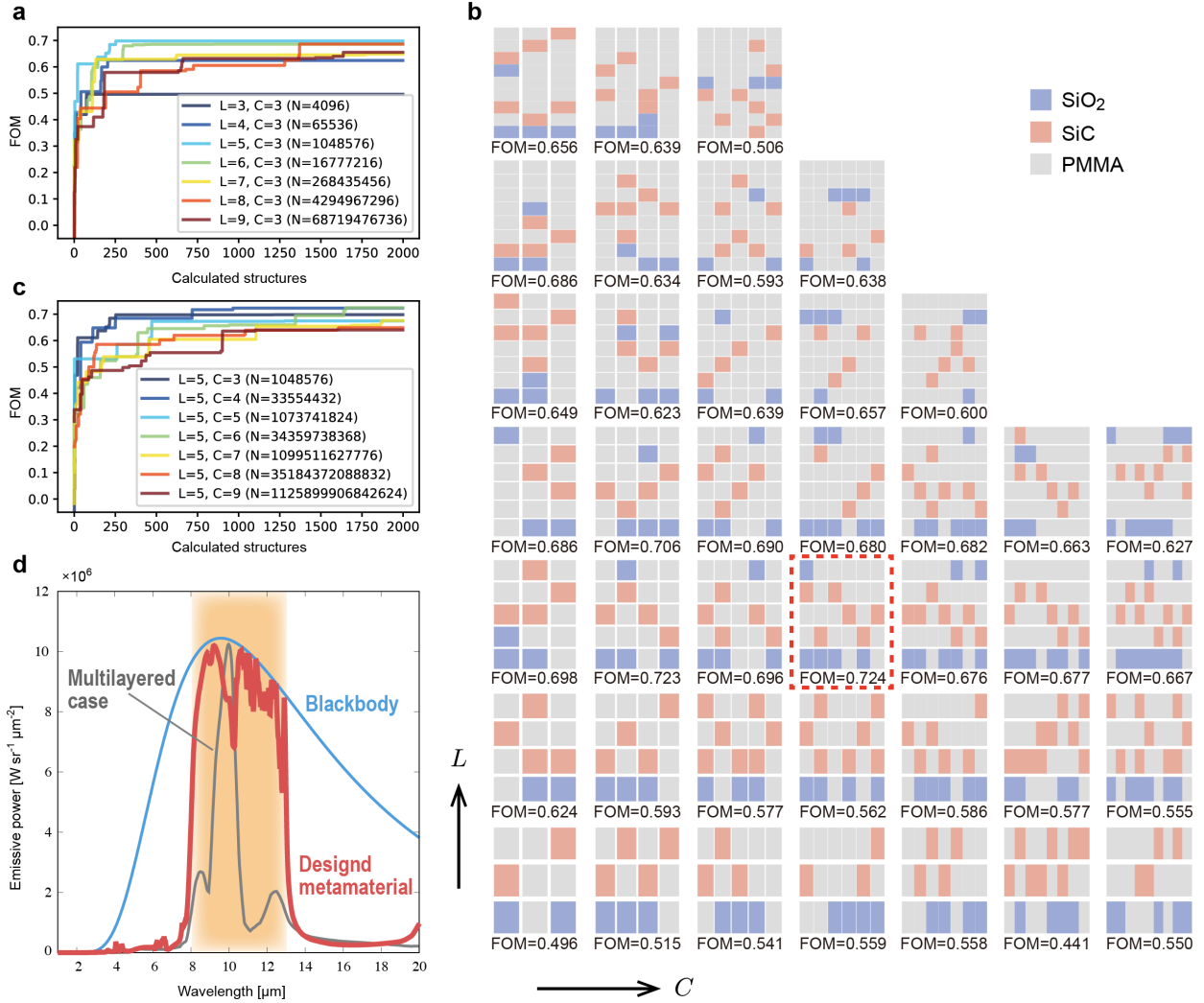

**Figure 4.** (a) Best FOM as a function of the iteration number for $C = 3$ with various $L$. (b) Structures with a high FOM designed by FMQA depending on $L$ and $C$. Dotted line denotes the structure with the highest FOM in our search. (c) Change in the best FOM for $L = 5$ with various $C$. (d) Emissive power of the designed structure with a highest FOM for the $L = 5$ and $C = 6$ case, which is the found best structure for various $L$ and $C$. For comparison, the multilayer optimum case with five layers is shown (see Supplementary Note B).

## Mechanism of high emittance in designed metamaterial

From the list of optimized structures, we noticed that structures with $SiO_2$ located separately at the top and bottom and with SiC mediated in the middle part always show a better FOM. To understand which part of the structure absorbs the wave energy, we evaluated the electric power dissipation density $w_e$ of each part. Figure 5 (a) shows $w_e$ for several typical wavelengths of the optimum structure designed in the above. The top and bottom $SiO_2$ absorb most of the wave energy within 8–11 μm, whereas the middle parts of the SiC layers dominate the absorption between 11 μm and 13 μm. Furthermore, around 11.8 μm, $SiO_2$ also facilitates in absorption.

Next, we addressed the mechanism of the high emittance of designed metamaterial. The emittance contour plot of the p-wave dispersion relation indicates that a high emittance is almost insensitive against the incident angle (see Supplementary Note C). Hence, the resonance is not due to the surface phonon polariton, which is highly sensitive to the incident angle. On the other hand, in terms of the magnetic polariton[51], diamagnetic response between the external field and centralized magnetic field inside the structure is usually excited. Consequently, it is almost insensitive of the incident angle. To further elucidate the magnetic polariton resonance, Fig. 5 (b) plots the magnetic field normalized by the maximum value at several typical wavelengths. Comparing Figs. 4 (d) and 5 (b) reveals that magnetic fields always shows a strong confinement at the part with a high emittance, which is where the polariton resonance is excited. Furthermore, when the emittance curve relatively falls, the confined magnetic field becomes flatter and less centralized. These results suggest that the high emittance of the designed structure originates from the magnetic polariton resonance. Since the multilayer structure with five layers only shows narrow absorption band between 8.5 μm and 10.5 μm (see Fig. 4 (d)), the structure design of materials is the key to realize the confinement of the magnetic field. To promote the understanding of our designed thermal radiator, the angle dependence and theoretical cooling power analysis are discussed in Supplemental Notes C and D.

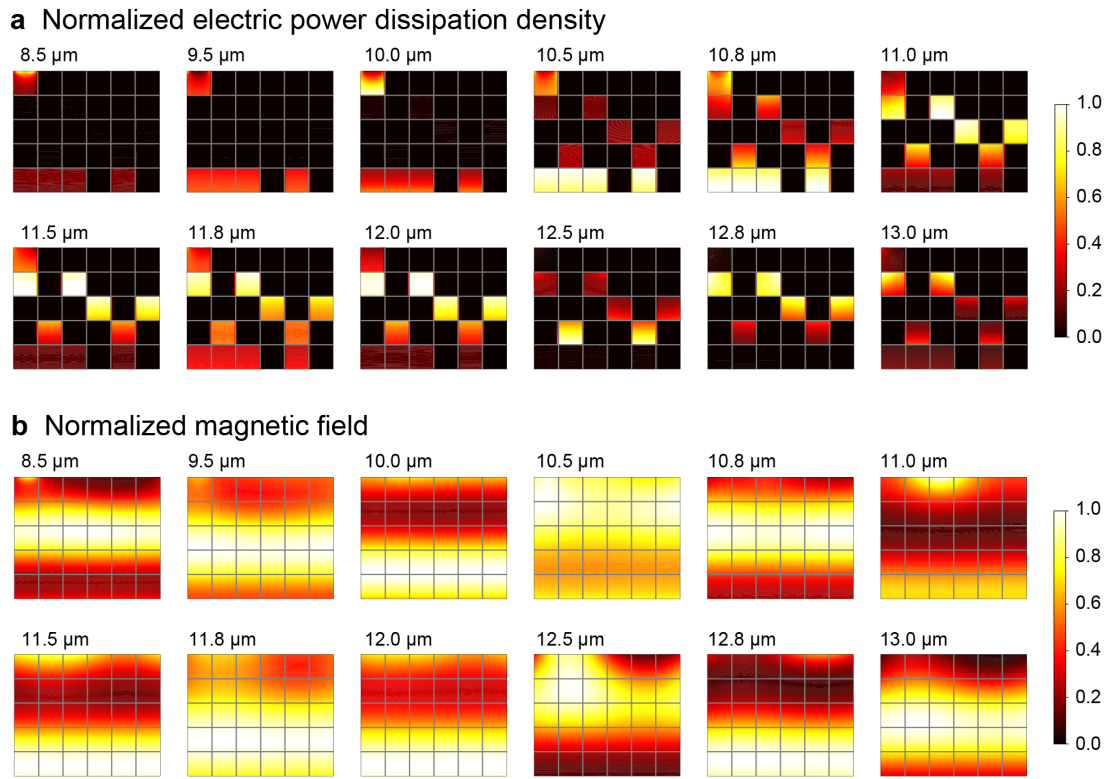

**Figure 5.** (a) Contour plot of the normalized electric power dissipation density at select wavelengths. (b) Contour plot of the normalized magnetic field.

## Discussion

In summary, we proposed a new optimization technique called FMQA, which uses a quantum annealer for the automated materials discovery. In our quantum-classical hybrid algorithm, the selection of the next candidate material with respect to acquisition function is represented as combinatorial optimization by using an FM, and this optimization problem is solved by the quantum annealer. By performing FMQA using the D-Wave 2000Q, we demonstrate that a metamaterial can be designed for radiative cooling within a small number of RCWA simulations. In the target metamaterial, which has $SiO_2$ and SiC wires placed in PMMA, a high FOM of 0.724 is achieved for radiative sky cooling. Compared with previous human-designed structures, the targeted single polarization FOM is far greater than the best reported structures, and the polarization- and angle-averaged cooling power is comparable (see Supplementary Note D). Although fabrication is beyond the scope of this paper, a stratified structure similar to the designed structures has been fabricated by the current technology[52], and thus, we believe that fabrication of the designed optimal structure is possible and it will be discussed elsewhere. Our algorithm is applicable to structural optimization of any properties as long as the property calculation is relatively fast with respect to the optimization process. Therefore, together with the rapid advance in computational physics and chemistry, the algorithm, in addition to straightforward extension such as tailoring spectral and angular-dependent radiative heat transfer, is expected to be used to control transport various carriers such as phonons, electrons, magnons to contribute to creation of new energy materials.

To the best of our knowledge, it is the first black-box optimization algorithm using an Ising machine such as a quantum annealer, while Ising machines are conventionally used to optimize explicitly-defined functions and to train machine learning models[24,25]. In future, the application domain of our algorithm will expand to even larger problems, as next-generation quantum annealers or other Ising machines equipped with many bits[53–56] become available. The computational bottleneck of our algorithm lies in the atomistic simulation conducted by a classical computer, but it will be resolved when materials simulations by quantum computing[57–64] are put into reality.

## Methods

### Simulation by RCWA

To calculate the thermal emissivity properties of the target metamaterials (Fig. 2), RCWA was employed. RCWA is semi-analytical method to solve Maxwell's equation and provides a high numerical accuracy (see Supplementary Note E). The spatial distribution of the dielectric constant and the involved electromagnetic field were decomposed by the $x$ and $z$ directions. By imposing boundary conditions in the $x$ and $z$ directions, the governing Maxwell equation can be solved quickly and accurately.

In our calculations, although PMMA was not universal in identical materials due to the influence of the fabrication process, we assumed that the PMMA was pure and the refractive index was fixed as 1.48 for simplicity[65]. The dielectric functions of $SiO_2$ and SiC were obtained from the tabulated data from Palik[66] with interpolation. Comparisons with the experimental results validated[67] that the RCWA method is a credible approach to design metamaterials for thermal radiators. Figure 2 shows a calculated example of the emissive power.

As a good thermal radiator for radiative cooling, the emittance spectra should fall within the wavelength region between 8 μm and 13 μm. To evaluate the likelihood that the designed metamaterial is the ideal case, the figure-of-merit (FOM) is defined as

$$\text{FOM} = \frac{\int_{\lambda_i}^{\lambda_f} \varepsilon_\lambda E_{b\lambda} \, d\lambda}{\int_{\lambda_i}^{\lambda_f} E_{b\lambda} \, d\lambda} - \frac{\int_{\lambda_{min}}^{\lambda_i} \varepsilon_\lambda E_{b\lambda} \, d\lambda}{\int_{\lambda_{min}}^{\lambda_i} E_{b\lambda} \, d\lambda} - \frac{\int_{\lambda_f}^{\lambda_{max}} \varepsilon_\lambda E_{b\lambda} \, d\lambda}{\int_{\lambda_f}^{\lambda_{max}} E_{b\lambda} \, d\lambda}, \tag{1}$$

where $\lambda_i = 8$ μm, $\lambda_f = 13$ μm, $\lambda_{min} = 1$ μm, and $\lambda_{max} = 20$ μm. Here, $\varepsilon_\lambda$ and $E_{b\lambda}$ are the spectral emittance and spectral blackbody emissive power calculated by RCWA, respectively. It should be noted that the spectral emittance is equal to the spectral absorptance in the thermal equilibrium state according Kirchoff's law. In the following calculation, the spectral emittance property was obtained directly from the spectral absorptance.

In addition, to quantify dissipated power absorbed by the structure, the electric power dissipation density $w_e$ is calculated as[68]

$$w_e = \frac{1}{2} \varepsilon_0 \varepsilon_{\mathrm{Im}} \omega |\mathbf{E}|^2,$$

(2)

where $\varepsilon_0$ is the permittivity in a vacuum, $\varepsilon_{\mathrm{Im}}$ is the imaginary part of dielectric function, $\omega$ is the angular frequency, and $\mathbf{E}$ is the complex electric field calculated by RCWA.

**Learning by factorization machine**

Using a D-Wave quantum annealer, the ground state of the quadratic unconstrained binary optimization (QUBO) can be effectively obtained with high speed and high accuracy. QUBO with $N$ bits is given by

$$H = \sum_{i=1}^{N} \sum_{j=1}^{N} Q_{ij} x_i x_j,$$

(3)

where $x_i$ is the 0/1 binary bit and $Q_{ij} = Q_{ji}$ takes real values. The key of our idea is to use a machine learning regression model, which can be expressed by QUBO, and the next candidate material structure with a high acquisition function can be rapidly selected by the quantum annealer. For the prediction, we utilized an FM, which is given by

$$y(\mathbf{x}) = \sum_{i=1}^{N} w_i x_i + \sum_{i=1}^{N} \sum_{j=1}^{N} \sum_{k=1}^{K} v_{ik} v_{jk} x_i x_j,$$

(4)

where $\mathbf{x} = \{x_1, ..., x_N\}$ determines the structure of the target metamaterial (see Fig. 2). In this model, as the size of factorization $K$ decreases, the number of fitting parameters $\{v_{ik}\}$ can be reduced. Note that the $K = N$ case is equivalent to the regression model by the QUBO itself. This should realize a good prediction without overfitting when the number of training data is small (see Supplementary Note F). Thus, this model should be suitable for automated materials discovery due to the small number of training data.

In this paper, the size of factorization was fixed as $K = 8$, which is the default parameter in the libFM package[69]. In the training, the negative FOM was inputted as $y(\mathbf{x})$, and some parameters ($\{w_i\}$ and $\{v_{ik}\}$) were tuned to predict the negative FOM with a higher accuracy. That is, the predicted FOM was used as the acquisition function itself. In this paper, these parameters were determined by Adam (adaptive moment estimation) for the training data set.

**Selection by the D-Wave quantum annealer**

We utilized the D-Wave 2000Q quantum annealer to select the next candidate material in automated materials discovery. The D-Wave 2000Q had 2038 qubits on the chimera graph. An FM was implemented on the fully connected graph, and the D-Wave 2000Q could create it with 63 nodes by regarding some qubits as one variable. For this embedding, we used the dwave.embedding.chimera.find_clique_embedding method in dwave-system[70]. Thus, in our quantum-classical hybrid algorithm for automated materials discovery, the maximum problem size was 63, but the fast selection from at most $2^{63}$ candidates is very attractive for materials science. The trained parameters ($\{w_i\}$ and $\{v_{ik}\}$) in the FM were converted to the parameters ($\{Q_{ij}\}$) in the QUBO format. As we set num_reads = 50, which is the parameter for the D-Wave 2000Q, 50 states were outputted as the ground state candidates within 16 ms as the QPU time. We selected the state with the minimum energy obtained by the D-Wave 2000Q from 50 states, and the material structure characterized by this state was chosen as the next candidate structure. After the candidate metamaterial structure was selected, we evaluated the FOM by RCWA and the number of training data for FM increased. By repeating this procedure (Fig. 1), the metamaterial structure with a high FOM can be obtained with small number of evaluations (simulations or experiments). So far, Ising machines including a quantum annealer were conventionally utilized for training in machine learning[24,25]. Hence, an application of the D-Wave's computer to the selection part, which corresponds to perform a lot of inferences, brings out a new ability of Ising machines.

Note that in our implementation, since the already observed structures were not excluded as the candidates in the selection by the D-Wave 2000Q, it was possible to select the already observed structures. When such a situation occurred, we randomly chose a metamaterial structure from the unobserved structures as the next candidate.

# Acknowledgements

We thank Kei Terayama, Masato Sumita, and Kotaro Tanahashi for the useful discussions. This article is based on the results obtained from a project subsidized by the "Materials Research by Information Integration" Initiative (MI2I) project and Core Research for Evolutional Science and Technology (CREST) (Grants No. JPMJCR1502, No. JPMJCR16Q5, and No. JPMJCR17J2) from the Japan Science and Technology Agency (JST). This work was supported partially by KAKENHI Grants (Grand No. 15K17720, No. 15H03699, and No. 16H04274) from the Japan Society for the Promotion of Science (JSPS). S.T. was also supported by JST PRESTO (Grant No. JPMJPR1665). K.T. was supported by a Grant-in-Aid for Scientific Research on Innovative Areas "Nano Informatics" [Grant 25106005] from JSPS.

# Author contributions

K.K. established the programing code for machine learning, performed computational experiments, and analyzed the data. J.G. established the programing code for materials simulations, performed the experiments, and analyzed the data. S.J. and S.T. provided critical suggestions on the direction of the present study and edited the manuscript. K.T, J.S., and R.T. conceived the project and wrote the manuscript. All authors discussed the results and implications, and provided comments on the manuscript. K.K. and J.G. contributed equally to this work.

# Competing interests

The authors declare no competing interests.

# Corresponding authors

Correspondence to Koji Tsuda (tsuda@k.u-tokyo.ac.jp), Junichiro Shiomi (shiomi@photon.t.u-tokyo.ac.jp), and Ryo Tamura (tamura.ryo@nims.go.jp).

# Supplementary information for
# Expanding the horizon of automated materials discovery
# via quantum annealing

**Koki Kitai[1], Jiang Guo[2], Shenghong Ju[2,3], Shu Tanaka[4,5], Koji Tsuda*[1,3,6], Junichiro Shiomi*[2,3,6], and Ryo Tamura*[1,3,6,7]**

[1]Graduate School of Frontier Sciences, The University of Tokyo, Chiba 277-8568, Japan

[2]Department of Mechanical Engineering, The University of Tokyo, Tokyo 113-8654, Japan

[3]Research and Services Division of Materials Data and Integrated System, National Institute for Materials Science, Ibaraki 305-0047, Japan

[4]Green Computing Systems Research Organization, Waseda University, Tokyo 162-0042, Japan

[5]JST, PRESTO, Saitama 332-0012, Japan

[6]RIKEN Center for Advanced Intelligence Project, Tokyo 103-0027, Japan

[7]International Center for Materials Nanoarchitectonics, National Institute for Materials Science, Ibaraki 305-0044, Japan

**Supplementary Table I.** Empirical computing time to perform 500 iterations in our automated materials discovery depending on the encoding bits. For the selection part, quantum annealer (quantum-classical hybrid) and an exhaustive search on a classical computer with an Intel Xeon E5-2690 v3 @ 2.6 GHz (only classical) are utilized. The learning by an FM and simulation by RCWA were conducted on the same classical computer.

| Bits | Quantum-classical hybrid | | | Only classical | | |
|------|-----------|----------|------------|-----------|----------|------------|
|      | Selection | Learning | Simulation | Selection | Learning | Simulation |
| 9    | 361 [sec] | 54.9 [sec] | 2920 [sec] | 236 [sec] | 50.0 [sec] | 2830 [sec] |
| 12   | 385 [sec] | 53.0 [sec] | 3620 [sec] | 1730 [sec] | 50.4 [sec] | 2630 [sec] |
| 15   | 423 [sec] | 51.1 [sec] | 3390 [sec] | 12500 [sec] | 45.0 [sec] | 2920 [sec] |
| 18   | 374 [sec] | 52.2 [sec] | 3140 [sec] | 110000 [sec] | 50.3 [sec] | 3170 [sec] |

**Supplementary Note A. Comparison between a global search and a local search**

To circumvent the difficulty of global optimization, a local search method is a powerful tool. In this supplementary note, we compared the efficiencies of our automated materials discovery when the local search method and the D-Wave quantum annealer are utilized to select the next candidate metamaterial. Here, the greedy algorithm is used to search the minimizer of the trained factorization machine (FM) as the local search method. In our greedy algorithm, the search space for the next sampling point in the acquisition function is restricted into the states that can be generated by a single bit flip from the present state. This constraint dramatically reduces the search space and achieves very high-speed search, but the optimization run could be caught in a local optimum.

In our implementation, one bit from a randomly generated initial state is iteratively flipped such that the acquisition function becomes smaller. If no bit can be updated, the final state is used as the next candidate metamaterial structure. Figure A-1 is the comparison plot for the best figure-of-merits (FOMs), which depend on the number of calculated structures using a global search (D-Wave quantum annealer) and a local search (greedy algorithm) for the $L = 4$ and $C = 3$ case. Although the target system is small, the efficiency using a global search is better than that using a local search. This difference should become larger as the target system size increases. Thus, the obtained result suggests that a global search, that is, the D-Wave quantum annealer, is strongly recommended for our metamaterial design to avoid being trapped in a local optimum.

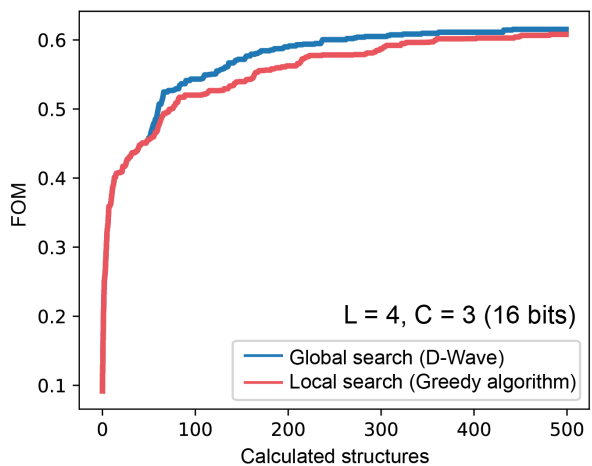

**Figure A-1.** Best FOM found during the optimization runs by a global search and a local search for the selection part averaged over 20 trials. First 50 states are randomly selected as common initial states. Target structure is size $L = 4$ and $C = 3$.

## Supplementary Note B. FOM in multilayer materials

We addressed the optimum structure of the multilayer metamaterials constructed by SiO$_2$, SiC, and PMMA. The optimum structure found in the main text has five layers (see Fig. 4 (b)). Hence, we searched for the five-layer structure for comparison where thickness of each layer is fixed as 1 μm and the periodic boundary condition was imposed for the horizontal direction. In this case, since the number of candidates is only $3^5 = 243$, all FOMs were calculated by RCWA. The top five structures with the highest FOMs are shown in Fig. B-1. The FOMs are much smaller than our designed thermal radiator (FOM = 0.724).

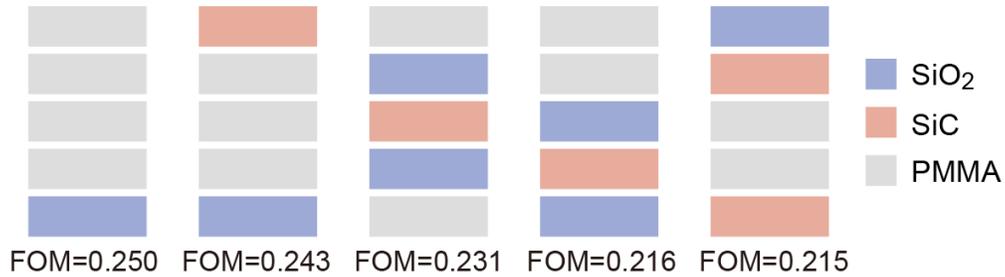

**Figure B-1.** Top five structures of the multilayer metamaterial with five layers.

## Supplementary Note C. Angle dependence of the designed thermal radiator

For an ideal selective radiator, emittance within the atmospheric window should be high for all solid angles (diffuse-like) in the hemisphere. We analyzed the dependence of the emittance on the polar angle and the azimuthal angle of the optimum structure found in the main text. Figures C-1 (a) and (b) show the polar angle $\theta$ dependence of emittance for p-polarized electromagnetic (TM) wave and s-polarized electromagnetic (TE) wave, respectively. The electric field of s-polarized wave and the magnetic field of p-polarized wave are along the $y$-axis. For the p-wave, the optimum structure keeps a very high emittance over a large range of angles up to nearly 75 degrees.

On the other hand, strong emittance within the atmospheric window is not always observed for the s-wave. In particular, for the region between 10.2 μm and 13.0 μm, the emittance becomes weaker, which might be originated from the intrinsic weak absorption of SiC in this wavelength range. This fact is expected because the anisotropic structure of metamaterial is targeted, and the phonon–polariton resonance along the $y$-axis can hardly be excited. For the azimuthal angle dependence property (Fig. C-1 (c)), selectivity between 8 μm and 13 μm is also obtained, but the magnitude of emittance falls rapidly after around 40 degrees.

Figure C-1 (d) shows the dependence of the average non-polarized incident emittance on the polar angle. The distinct selectivity between 8 μm and 13 μm is realized in a range of incident polar angles, which is also beneficial for the radiative cooling. While the performance is still limited by the above-discussed features of the s-wave, the fact we are able to achieve the high FOM for the p-wave in a large span of the incident polar angle (Fig. C-1(a)) indicates possibility to realize the omnidirectionally high FOM by extending the current two-dimensional design to three-dimensional for instance with the wires cross aligned in the $x$ or $y$ directions.

In Fig. C-2, the emittance contour under p-wave incidence is plotted as a function of the $x$-component of the wave vector $k_x$ and the angular frequency $\omega$. Here, the incidence angle is characterized by $\theta = \sin^{-1}(k_x c/\omega)$ where $c$ is the speed of light in a vacuum. The p-wave emittance contour dispersion relation is almost independent of the angle between 769 and 1250 cm$^{-1}$ (8–13 μm), which indicates the magnetic polariton resonance is excited in this wavenumber range. However, the surface phonon polariton resonance is also excited by relatively weak magnitude as can be seen from the surface phonon polariton dispersion line. In addition, coupling resonance between the surface phonon polariton and the magnetic polariton can be identified in this figure.

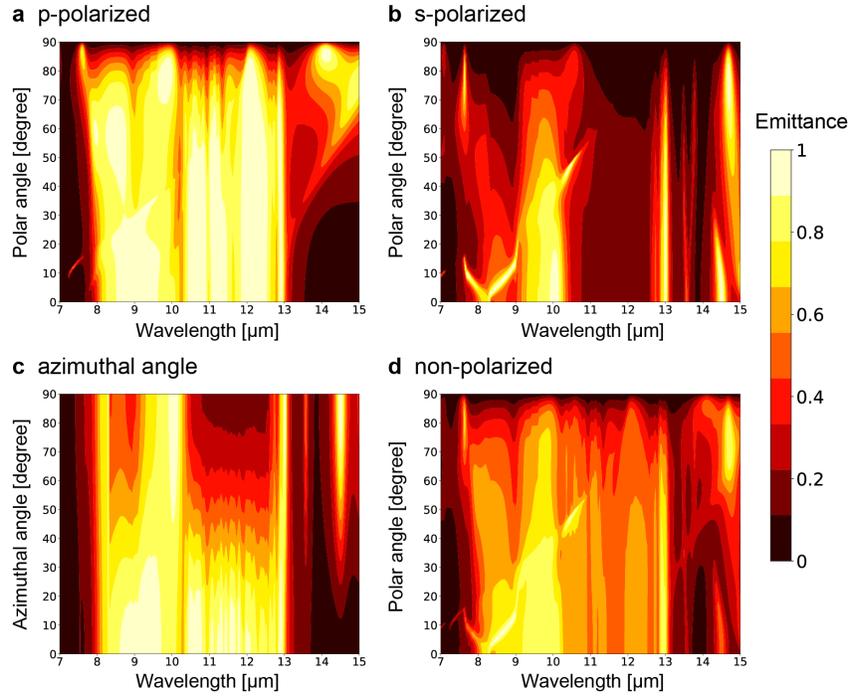

**Figure C-1.** Polar angle dependence of the emittance in the designed radiator for (a) the p-polarized electromagnetic wave and (b) the s-polarized electromagnetic wave. (c) Azimuthal angle dependence of the emittance, the polar angle is normal ($\theta = 0°$). (d) Polar angle dependence for the averaged non-polarized incidence.

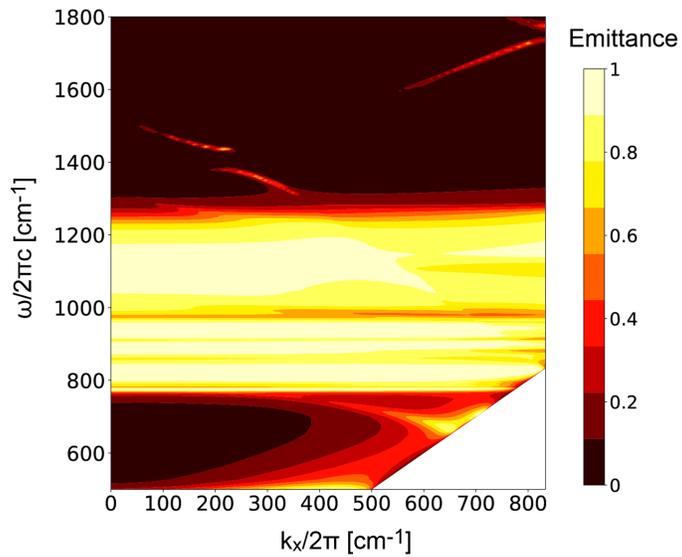

**Figure C-2.** Emittance dispersion plot of p-wave incidence.

**Supplementary Note D. Comparison of performance with previous works**

Firstly, we compare the obtained wavelength selective property with those of previous works on radiative cooling in terms of the FOM defined by equation (1) in the main text. Figure D-1 shows the comparison for emittance[1,2]. Our designed metamaterial shows the highest FOM among these radiators, which demonstrates that the designed one is closest to the ideal radiator in the view of wavelength selectivity aspect. Note that all of the emittance data taken from references are under normal incidence, and our emittance data is only for p-polarized incidence.

We further evaluate the performance in terms of the radiative cooling power, by taking the averaged non-polarized emissivity properties of our designed thermal radiator. We consider a representative case where thermal radiator with temperature $T_s$ is placed under an ambient temperature $T_a = 303$ K and is exposed to clear sky. Now, the outgoing energy is the radiated heat by the thermal radiator $P_{rad}(T_s)$, while the incoming energy includes downward atmospheric radiation $P_{atm}(T_a)$, directly absorbed solar radiation $P_{sun}$, and the conduction and convection heat transfer between the thermal radiator and ambient $P_{con}(T_s, T_a)$. The definitions of each power used here are explained in Refs. 1 and 3. Then, the net cooling power of thermal radiator $P_{net}(T_s, T_a)$ is defined as,

$$P_{net}(T_s, T_a) = P_{rad}(T_s) - P_{atm}(T_a) - P_{sun} - P_{con}(T_s, T_a). \tag{d.1}$$

For the absorbed atmospheric radiation, we used the MODTRAN software and calculated the atmospheric transmittance by the 1976 US standard atmosphere model[4]. For radiative cooling, a large positive net cooling power is prefered. In Fig. D-2, we plotted the normalized emitted radiation and absorbed atmospheric radiation as functions of the wavelength for the normal polar angle and p-wave case when $T_s = T_a = 303$ K. The absorbed atmospheric radiation is relatively small compare with the energy emitted out by the designed thermal radiator. Thus, a large positive net cooling power is expected.

Another important factor is the low equilibrium temperature of the radiator at which the outgoing and incoming energies are balanced, because the equilibrium temperature is directly related to the cooling of the radiator itself. We first considered the thermal conduction and convection coupled effect, that is, $P_{con}(T_s, T_a)$. Ref. 2 experimentally demonstrated that this effect can be controlled by insulation with plastic foam and covering with a shield. This conduction and convection heat transfer can be expressed as

$$P_{con}(T_s, T_a) = h_{cond+conv}(T_a - T_s), \tag{d.2}$$

where $h_{cond+conv}$ is the non-radiative heat coefficient. First of all, an ideal emitter for radiative cooling, which only emits and absorbs within the atmospheric window, is calculated for comparison. As depicted in Fig. D-3 (a), the ideal emitter possesses a cooling power of 100.36 W m$^{-2}$ at ambient temperature ($T_a = 303$ K) when $h_{cond+conv} = 0$. In this case, the equilibrium temperature is more than 50 K lower than the ambient temperature. On the other hand, our designed structure shows a cooling power of 60.62 W m$^{-2}$ when $T_s = T_a$ and $h_{cond+conv} = 0$. The differences between the equilibrium temperature of the radiator and the ambient temperature are 37.66, 22.09, 12.55, and 6.90 K when $h_{cond+conv}$ is equal to 0, 1, 3, and 6.9 W m$^{-2}$ K$^{-1}$, respectively.

Next, the solar energy absorption $P_{sun}$ is considered. The average solar irradiance intensity (950 W m$^{-2}$) is much larger than the maximum radiative cooling power (100.36 W m$^{-2}$), and thus the solar energy must be considered. Usually, solar reflectors[2,5] or integrating a solar reflecting thin film with the photonic thermal radiative cooling emitter[1,6] can reduce the total solar energy absorption to less than 3%. Therefore, we theoretically assumed that the solar absorption is 3% of the average solar irradiance intensity, and evaluated its influence on the cooling performance. As can be seen in Fig. D-3 (b), when 3% solar energy is added, the cooling power becomes a 32.12 W m$^{-2}$, and the equilibrium temperature difference is 18.05, 11.31, 6.58, and 3.65K.

Finally, we present the comparison of the temperature-dependent cooling power at the ambient temperature set to 293 K with the reference papers in Fig. D-4. Our designed device shows 42 W m$^{-2}$ cooling power at 293 K and achieves stagnation temperature at nearly 253 K. The cooling power of our design is smaller than Ref. 2 (62 W m$^{-2}$) due to the low emittance value for s-wave incidence but is larger than Ref. 1 (38 W m$^{-2}$). In terms of the temperature difference, our design reaches nearly 40 K, which is comparable to 48 K in Ref. 2, gaining from the high wavelength selectivity in 8–13 μm which substantially decreases the rate of slope of the cooling power. Therefore, the expected cooling power of the designed wire-based metamaterial is comparable to the best ones reported thus far.

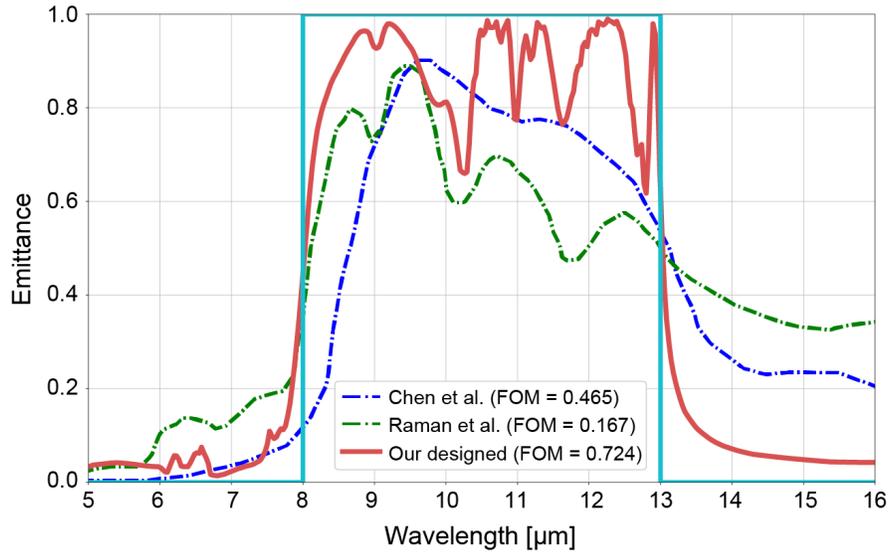

**Figure D-1.** Emittance of some radiators reported in previous works[1,2] and our designed radiator. Sky blue solid line indicates the ideal emittance for radiative cooling.

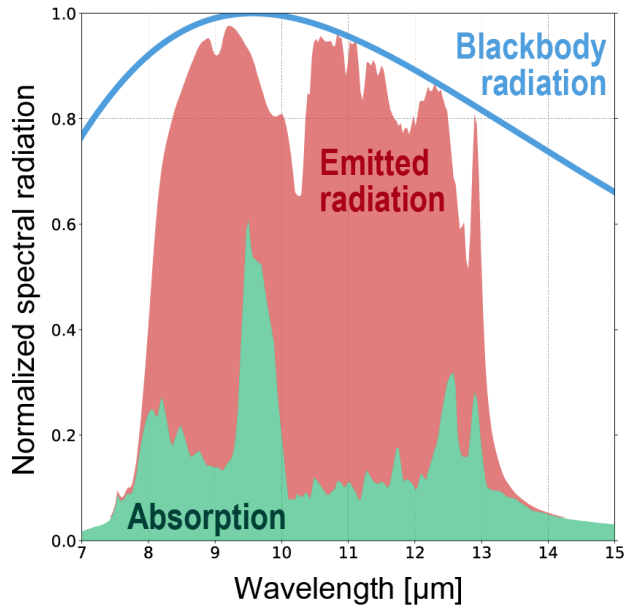

**Figure D-2.** Normalized radiative energy exchange of the designed structure. Red shading area: normalized spectral emitted radiation by the designed structure; Green shading area: normalized spectral absorption of the atmospheric radiation; Blue line: normalized blackbody radiation at 303 K.

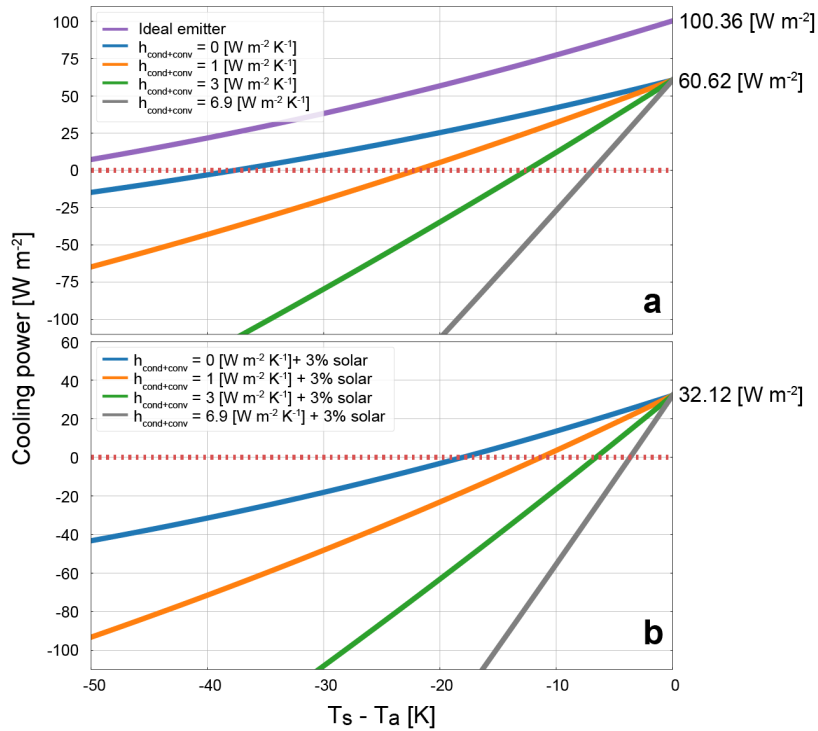

**Figure D-3.** Cooling power versus temperature difference between the thermal radiator and ambient (303 K) for various conduction and convection heat transfers. (a) Without solar radiation. (b) With 3% solar radiation. Radiator temperature $T_s$ at which the cooling power is equal to 0 is the equilibrium temperature.

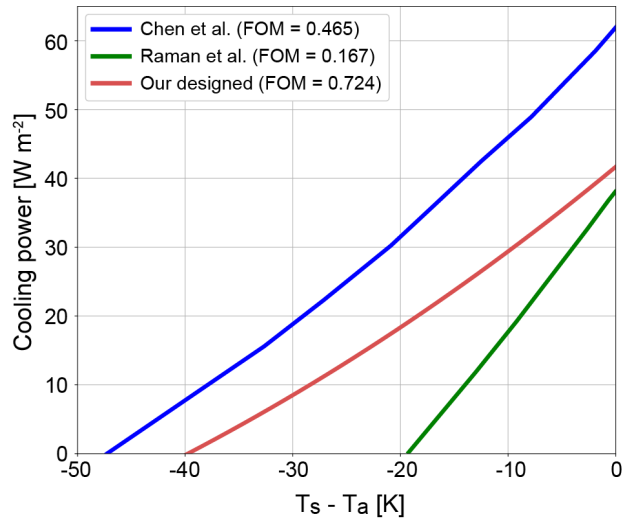

**Figure D-4.** Comparison of the temperature dependance of cooling power between our design and previous works assuming absence of parasitic heat load, the ambient temperature is set to 20ºC here in accordance to the reference paper[1,2].

## Supplementary Note E. Validation of RCWA calculation

Rigorous coupled wave analysis (RCWA)[1] is a semi-analytical method for fast computing diffraction efficiency of sub-wavelength periodic structures. A homemade RCWA parallel computing code is developed based on Python. To validate the correctness of our code, we compared the result from our code to that by the finite-difference time-domain (FDTD) method[2]. Figure E-1, which shows the results of the target structure treated in Ref. 3, agrees relatively well. Furthermore, we calculated the emissivity properties by the RCWA algorithm software $S^4$ [4,5], which reproduces exactly the same result with our code. These results indicate that our code is credible.

It is very difficult to predict how many harmonic Fourier terms $N_{Fourier}$ are necessary to obtain a convergent result at the beginning time. When $N_{Fourier}$ is too large, the simulation time increases and the memory is at unacceptable level. Thus, there is a tradeoff between the accuracy and simulation time. To understand the appropriate value of $N_{Fourier}$, we calculated the emissivity properties depending on $N_{Fourier}$ for 200,000 randomly generated metamaterial structures. Here, the target metamaterial structure is the $L = 6$ and $C = 9$ case, which is the largest size searched in the main text. Figure E-2 is the frequency histogram of $N_{Fourier}$ required for convergence. For most of structures (over 99%), $N_{Fourier} = 40$ is sufficient for convergence. Furthermore, the difference in FOMs between the convergent result and the result only by $N_{Fourier} = 40$ is no more than 0.05 in all cases. Thus, to save the simulation time, we fixed the number of Fourier terms to 40 in our FOM calculations.

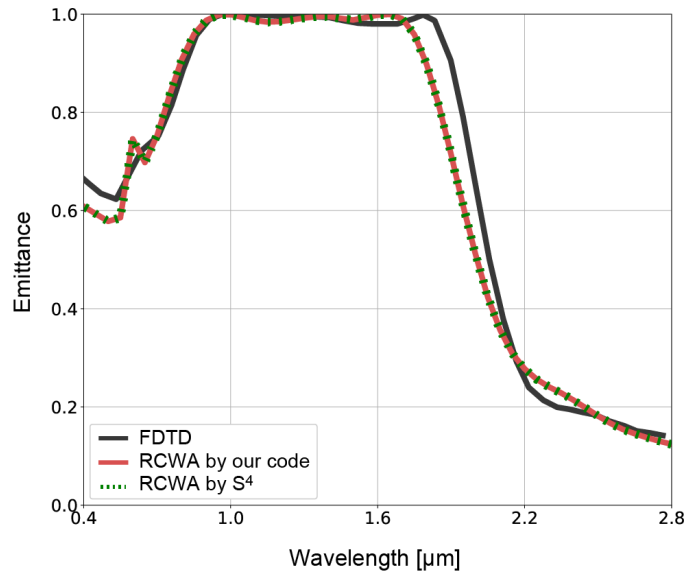

**Figure E-1**. Emittance calculated by FDTD, our RCWA code, and $S^4$ for the structure treated in Ref. 3.

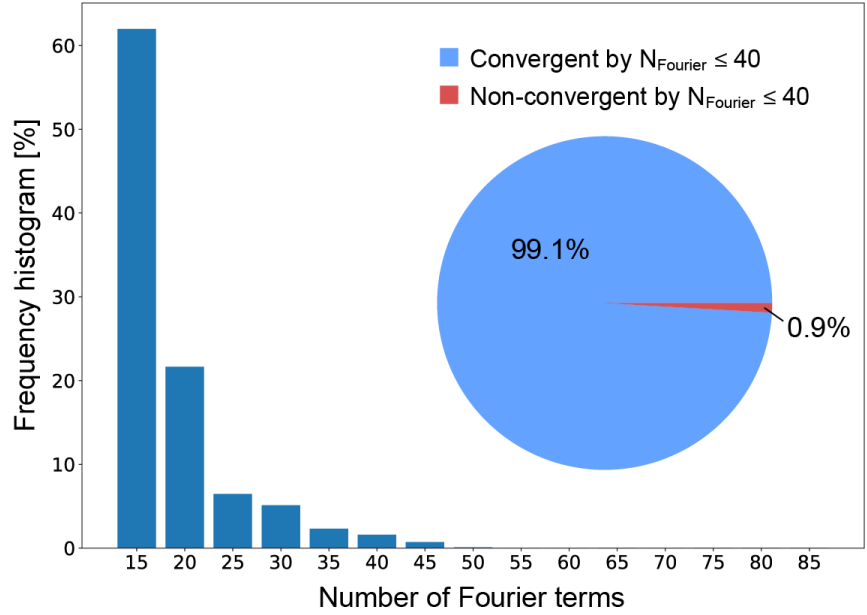

**Figure E-2**. Frequency histogram of the number of Fourier terms $N_{Fourier}$ required for convergence of 200,000 structures. For the 99.1 % structures, $N_{Fourier} = 40$ is sufficient to obtain the convergent result.

**Supplementary Note F. Dependence of the factorization machine on the size of factorization**

To avoid overfitting, the size of factorization $K$ in an FM must be appropriately determined. In this supplementary note, we consider the $K$ dependence of the prediction accuracy for the $L = 4$ and $C = 3$ case. Figure F-1 is the five-fold cross-validation error, which determines the typical regression errors in the prediction with $K$ for a various number of training data. The training dataset is prepared by a random sampling in all states, and the average of ten independent calculations with different training data selections was evaluated. We observed that the cases with fewer factors exhibit a higher accuracy when the number of training data is relatively small ($< 200$). This implies that for larger $K$, the number of fitting parameters is too large against the training data size, resulting in overfitting. Hence, the reduction of the size of factorization is essential to keep a high regression accuracy when the number of training data is small. For simplicity of implementation, we fixed $K = 8$ for all optimization runs, which is the default parameter in the libFM package. In the future, we will develop a method to decide an appropriate value of $K$ based on the training data for automated materials discovery.

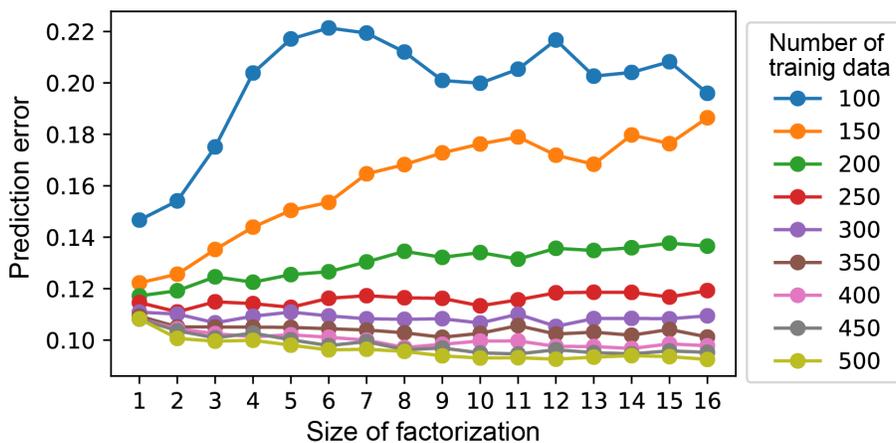

**Figure F-1**. Change in the prediction errors as a function of the size of factorization $K$ in an FM for various amounts of training data. Prediction errors are evaluated by RMSD by performing a five-fold cross validation.